\documentclass[aip,reprint]{revtex4-1}
\usepackage{dcolumn}
\usepackage{color}
\usepackage{amsmath}
\usepackage{amssymb}

\usepackage[dvips]{graphicx}

\newcommand{\be}{\begin{equation}}
\newcommand{\ee}{\end{equation}}
\newcommand{\bea}{\begin{eqnarray}}
\newcommand{\eea}{\end{eqnarray}}

\newcounter{Fig}

\begin{document}

\begin{sloppy}

\title{Giant terahertz near-field enhancement by two-dimensional plasmons}

\author{Arthur R. Davoyan} \email{arthur.davoyan@gmail.com}
\author{Vyacheslav V. Popov}
\affiliation{Kotelnikov Institute of Radio Engineering and Electronics (Saratov Branch), Russian Academy of Sciences, Zelenaya 38, Saratov 410019, Russia}
\affiliation{Saratov State University, Astrakhanskaya 83, Saratov, 410012, Russia}
\author{Sergey A. Nikitov}
\affiliation{Saratov State University, Astrakhanskaya 83, Saratov, 410012, Russia}
\affiliation{Kotelnikov Institute of Radio Engineering and Electronics, Russian Academy of Sciences, Mokhovaya 11-7, Moscow, 125009, Russia}
\date{\today}

\begin{abstract}

We consider a periodically gated two-dimensional electron system, with a central gate finger biased independently creating a tunable plasmonic cavity in a planar plasmonic crystal. We demonstrate that the plasmons resonantly excited in the periodic plasmonic lattice by incident terahertz radiation can strongly pump the cavity plasmon modes leading to a deep sub-wavelength concentration of terahertz energy with giant electric field enhancement factor up to $10^4$.
\end{abstract}

\maketitle

Many material properties are sensitive to terahertz (THz) radiation, which makes THz spectroscopy an essential tool for the analysis of material microstructure, chemical composition and for the medical treatment~\cite{THz_book,THz_nature}. However, the diffraction limit of conventional optical systems, i.e. their inability to focus light in deep subwavelength regime, challenges the study of individual micro- and nanostructures.

Recent works have extensively discussed concentration of the THz energy beyond the diffraction limit with metallic structures, including, nanoapertures~\cite{Nahata}, tapers~\cite{Stockman,Taper,Mittleman} and metal edges~\cite{THz_focus}. In particular, in Refs.~\cite{Stockman,Taper} it was shown that tapered metallic rods and wedges can focus THz and mid-infrared radiation into a nanometer-size spots. In Ref.~\cite{Kim}, THz energy concentration in a metallic nanoslit beyond the skin-depth limit was demonstrated with field enhancement factors up-to $10^3$ at $0.1$ THz, nevertheless, the effect was steeply decreasing with frequency increase (with factors below $100$ at $1$ THz). One of the key approaches for the design and study of the metallic nanostructures for THz photonics is making parallels with optical plasmonics. However, at THz frequencies the electromagnetic field is weekly localized near the metal surface and can be confined only between closely placed metallic surfaced or in dispersion-engineered geometries~\cite{Spoof1,Jinho}. Efficient coupling of THz radiation into the metallic nanocomponents demands comprehensive antenna design, which, together with the lack in tunability, makes the development of functional integrated THz photonic devices based on metallic nanostructures a challenging task.

Employing structures supporting intrinsic {\it two-dimensional (2D) plasmons}, i.e. waves of free electron density physically localized in a plane, at THz frequencies, may resolve these problems and change significantly the design principles of THz photonic devices. Among such structures are semiconductor quantum wells, supporting 2D electron gases, and recently discovered graphene~\cite{Graphene}, a natural 2D material. The potential prospectives of graphene for THz photonics are currently an active research field \cite{Graphene_nature,Graphene_Abajo,Graphene_Engheta}. Contrary to metal structures, the properties of such materials can be easily tuned by the applied electric potential~\cite{Graphene}. The latter effect was utilized for the development of compact tunable THz sources and detectors based on the field-effect transistor (FET) structures~\cite{Knap}. In Refs.~\cite{Popov_obzor,Shaner_v1} the enhanced plasmonic response in a periodic grating gate FET was demonstrated, and in Ref.~\cite{Shaner_v2} a substantional increase in responsivity of such detector combined with a micro-bolometer element created by a potential barrier via splitting grating-gate was achieved. However, to the best of our knowledge, the THz field concentration and its enhancement in such 2D plasmonic structures has not been discussed yet.


\begin{figure}[h]
  \begin{center}
      \includegraphics[width=1\columnwidth]{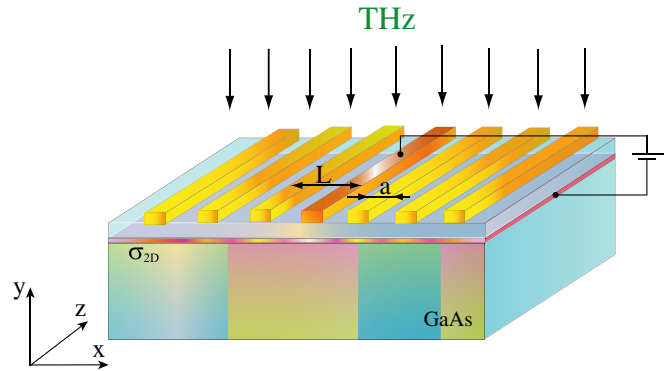}
      \caption{(Color online) Schematic of the split-grating-gate field effect transistor, where the central gate contact is biased individually.}\label{scheme}
  \end{center}
\end{figure}

In this letter, we propose a novel mechanism for {\it tunable} THz near-field enhancement at a defect in a planar plasmonic crystal. We study plasmon excitation in the FET structure with a split-grating gate, where the central gate finger is biased independently from the rest of the grating gate. We show that such configuration creates a tunable plasmonic cavity in the a planar plasmonic crystal lattice. We analyze the excitation of the lattice and the cavity plasmons, and demonstrate that their interaction leads to a dramatic THz near-field enhancement.

We assume that a THz plane wave, with electric field perpendicular to the grating gate fingers, ${\bf E}\, || \, {\bf x}$, is incident normally upon the structure as shown schematically in a Fig.~\ref{scheme}. A periodic metallic grating with period $L=4 \mu$m ($2 \mu$m metal, $2 \mu$m gap) is placed on the top of GaAs substrate with dielectric permittivity $\varepsilon=12.8$. We suppose that $200$nm below the wafer surface a 2D electron gas with the electron density $N_0=2.5\times 10^{11}$cm$^{-2}$ is formed.

\begin{figure}[h]
  \begin{center}
      \includegraphics[width=1\columnwidth]{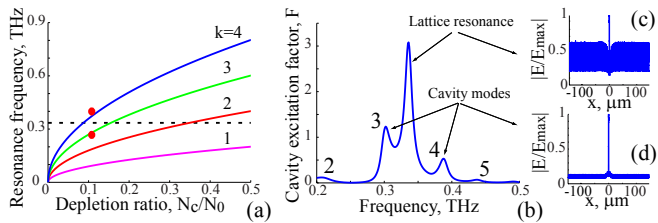}
      \caption{(Color online)(a) Plasmonic cavity resonance frequency as a function of 2D electron density under the defect gate for four cavity modes. Dashed line shows the position of the plasmonic lattice resonance. Dots mark the positions of
      the 3-rd and 4-th order cavity resonances shown in panel (b). (b) Cavity excitation factor spectrum for the cavity depletion ratio $N_c/N_0 = 0.12$. Numbers indicate the resonance orders. (c),(d) The mode profiles at the lattice resonance ($\omega = 0.336$ THz) and at the 4th order cavity resonance ($\omega = 0.388$ THz), respectively.}\label{modes}
  \end{center}
\end{figure}

Negative electric potential applied to the central gate finger depletes the 2D electron channel under this particular finger. Therefore, the biased area acts as a tunable plasmonic cavity (defect) in a periodic plasmonic lattice. Note that, the grating gate, if biased, also creates periodic modulation of electron concentration in the 2D electron channel, and hence the frequency of the plasmon resonance of the plasmonic lattice can be also changed. Here, without loss of generality, we consider that the grating gate is not biased and the 2D electron density in the grating gated regions of the channel, $N_0$, remains homogeneous.

When the frequency of the incident THz wave, $\omega$, is far away from the lattice resonances, the lattice plasmons can not be excited and the 2D channel response is governed mainly by the Drude background conductivity. In this case, neglecting the excitation of the lattice plasmons, the plasmonic cavity can be fairly well modeled by a homogeneous 2D electron system gated only by a single gate finger.
This system was analyzed in details in Ref.~\cite{Polischuk}, where it was shown that the cavity resonance frequencies roughly satisfy a simple electrostatic formula:

\begin{equation}\label{cavity}
  \omega_{def}^{(k)} = \sqrt{\frac{N_ce^2h}{m^*\varepsilon_0\varepsilon}}\frac{\pi k}{a+2h},
\end{equation}
where $k$ is the number of the resonance order, $N_c$ is 2D electron density in the channel under the cavity gate, $h$ is the gate-to-channel separation, and $a$ is the width of the gate finger.

\begin{figure}[h]
  \begin{center}
      \includegraphics[width=1\columnwidth]{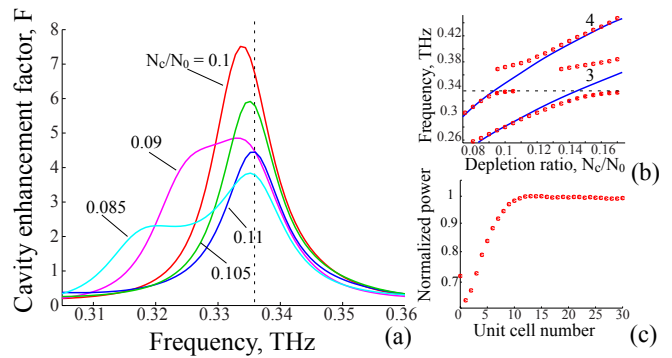}
      \caption{(Color online)(a) Cavity excitation factor spectrum for different depletion ratios, $N_c/N_0$. Dashed line shows the position of the plasmonic lattice resonance. (b) Dispersions of the 3-rd and 4-th cavity resonances as functions of the the cavity depletion ratio. Solid curves show the dispersion of corresponding unperturbed cavity modes. Dashed straight line marks the position of the plasmonic lattice resonance. (c) The power stored in the unit cell of the plasmonic lattice with increasing the lateral distance from the cavity. (enhanced online)}\label{res}
  \end{center}
\end{figure}

In Fig.~\ref{modes}(a), we plot the dispersion of the cavity resonance frequency defined by Eq.(\ref{cavity}) for four cavity modes with the variation of electron density in the biased area of the channel. In the same figure, we mark the position of the fundamental resonance of the plasmonic lattice. According to Fig.~\ref{modes}(a) the dispersions of the higher-order cavity resonances cross the plasmonic lattice resonance dispersion with depletion of the cavity channel. In this case, we expect strong interaction between the cavity and the lattice plasmon modes.

To prove our predictions, we perform comprehensive numerical simulations using the finite element method. We perform 2-D simulations, considering no variation of the field in the ${\bf z}-$direction and searching for the field profiles in the (\textit{x-y}) plane. We approximate the grating gate by the sequence of infinitely thin perfect electric conductors. To model the structure, we use the supercell approximation Ref.~\cite{Fan}. In this method, we embed our structure of interest into a large supercell periodically repeated in the ${\bf x}-$direction. We consider that the supercell contains the defect and 36 periods of the grating gate on either side of it. The electromagnetic response of 2D electron gas in a local approximation can be described by the 2D Drude conductivity, $\sigma_{2D}(\omega) = N(x) e^2 \tau/m^*(1-j\omega\tau)$, where $N(x)$ is the 2D electron density ($N(x)=N_c$ in the cavity and $N(x)=N_0$ in the rest of the channel), $\tau$ is the electron scattering time.

We introduce the cavity excitation factor as a ratio between the THz power absorbed in the cavity and the energy flux in the incident THz wave, $F=\frac{1}{2a}Re (\int_{defect} \sigma |E_x|^2 dx) /P_{in}$. First, we demonstrate the validity of our approach for the prediction of the cavity resonance positions. Using Eq.~(\ref{cavity}) and Fig.~\ref{modes}(a), we choose the the depletion ratio $N_c/N_0 =0.12$ in such a way that the 3-rd and the 4-th cavity resonances are relatively close to the plasmonic lattice resonance. We plot spectrum of the cavity excitation factor for this case in Fig.~\ref{modes}(b). To identify the excited plasmon modes at each resonance, we analyze corresponding electric field profiles, see Fig.~\ref{modes}(c,d). The analytically predicted positions of the lattice and the cavity resonances reasonably well coincide with the results of numerical simulations, see Fig.~\ref{modes}(a). The resonances of the cavity modes with the eigen-frequencies close to the plasmonic-lattice-resonance frequency become stronger due to the strong intermode interaction; whereas the cavity modes are negligibly weak far away from the lattice plasmon resonance.

\begin{figure}[t]
  \begin{center}
      \includegraphics[width=1\columnwidth]{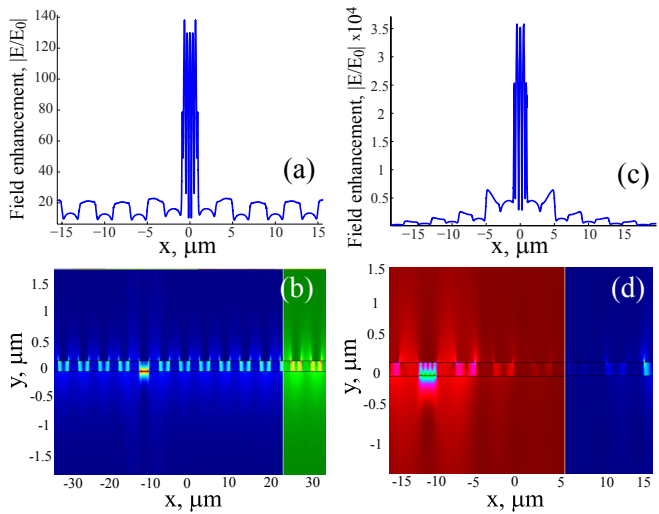}
      \caption{(Color online)(a) Normalized electric-field profile in the 2D electron channel and (b) maps of the electric-field distribution in the (\textit{x-y}) plane for $\tau = 10^{-11}$s; (c),(d) the same as in (a) and (b) for $\tau = 10^{-10}$s}\label{enh}
  \end{center}
\end{figure}

Moving further, we study the interaction between the lattice and the cavity plasmon modes. The cavity plasmon resonances are shifted to lower frequencies for decreasing depletion ratio, see Fig.~\ref{modes}. For depletion ratio $N_c/N_0 \simeq 0.1$ the the 4-th order cavity resonance merges with the lattice plasmon resonance leading to the increase of the resonance amplitude, see Fig.~\ref{res}(a) and corresponding media. For even smaller $N_c/N_0$, the resonances are separated again decreasing in amplitude, until the the 5-th cavity resonance approaches the lattice eigen-frequency. In Fig.~\ref{res}(b), we plot the positions of the cavity resonances as functions of the depletion ratio showing the interaction between the cavity plasmon modes and the lattice plasmon mode. When the cavity and the lattice plasmon resonances approach each other, they experience strong interaction leading to the {\it avoided crossing} of corresponding resonances.

In Fig.~\ref{enh}(a) and (b), we plot the electric-field enhancement factor, $|{\bf E}/{\bf E}_0|$ and the map of the electric-field distribution in the (\textit{x-y}) plane in the avoided crossing regime for $N_c/N_0 = 0.1$ and $\omega = 0.332$ THz. The electric field is greatly enhanced in the cavity up-to $150$. The plasmon excitations in the lattice unit cells adjacent to the cavity are suppressed. With increasing the lateral distance from the cavity, the plasmon energy in the lattice cell saturates at the level corresponding to the lattice plasmon resonance, see Fig~\ref{res}(c). This is due to the fact that the cavity mode is strongly pumped by the resonantly excited lattice plasmons. The $Q$-factor of the resonances is about $20\div30$ for a room temperature electron scattering time $\tau=10^{-11}$s. We have also studied THz near-field enhancement for longer electron scattering time $\tau = 10^{-10}$s (electron mobility $\mu \sim 10^6$cm$^2/$V)corresponding to cryogenic temperatures. In this case, the $Q$-factor is much higher up to $3000$ and hence the interaction between the cavity and the lattice plasmon modes is much stronger leading to a dramatic THz near-field enhancement factor $|{\bf E}/{\bf E}_0| \simeq 10000$, see Figs.~\ref{enh}(c-d), which, to the best of our knowledge, is much greater than any of previously reported values~\cite{THz_focus,Kim}. Our calculations show that such dramatic enhancement of THz near field by 2D plasmons can be achieved at higher THz frequencies (up to several THz) in the structure with a smaller grating-gate period and a sub-micron plasmonic cavity.

In conclusion, we have studied THz field enhancement in the planar plasmonic crystal formed by a splt-grating-gate FET structure with a defect induced by an individually biased central finger. We have shown that the interaction between the cavity and the lattice plasmon modes can be electrically tuned by the voltage applied to the central gate finger. We predict unprecedented THz electric field enhancement factor up-to 10000 with a deep subwavelengh concentration of the THz electric field. We believe that such strong THz electric field enhancement can be utilized for THz nonlinear applications and for the design of the next generation of THz plasmonic devices.

We thank Dr. A.E. Miroshnichenko and Dr. O.V. Polischuk for useful discussions. This work has been supported by the Russian Foundation for Basic Research (Grant Nos.10-02-93120 and 11-02-92101), by the Russian Academy of Sciences Program ``Fundamentals of Nanotechnology and Nanomaterials'', and by the Grant of the Government of the Russian Federation for supporting scientific research projects supervised by leading scientists at Russian institutions of higher education (Contract No. 11.G34.31.0030).

\end{sloppy}
\end{document}